\documentclass[preprint,floatfix,eqsecnum,aps,nofootinbib]{revtex4}

\usepackage{amsmath}
\usepackage{amssymb}
\usepackage{amscd}
\usepackage{mathrsfs}
\usepackage{graphicx}
\usepackage{subfigure}
\usepackage{slashed}
\usepackage{mathbbol}
\usepackage{subfigure}

\newcommand{\diag}{\mathop{\mathrm{diag}}}

\def\LE{\ensuremath{\ell_{E}}}  
\def\LP{\ensuremath{\ell_{P}}}  
\def\VL{\ensuremath{V_{\lambda}}}  
\def\Dsup{\ensuremath{D_\text{sup}}}  
\def\diag#1{\ensuremath{\left[\text{diag-}#1\right]}}  
\def\dim#1{\ensuremath{\left[#1\right]}}  

\begin{document}

\title{Lifshitz theories with extra dimensions and $3+1$-d Lorentz invariance}
\author{Xiao-Gang He}\email{hexg@phys.ntu.edu.tw}
\affiliation{INPAC, Department of Physics, Shanghai Jiao Tong University, Shanghai, China}
\affiliation{Department of Physics andÊ Center for Theoretical Sciences, National Taiwan University, Taipei, Taiwan}
\author{Sandy S. C. Law}\email{slaw@cycu.edu.tw}
\affiliation{Department of Physics, Chung Yuan Christian University, Chung-Li 320, Taiwan}
\author{Raymond R. Volkas}\email{raymondv@unimelb.edu.au}
\affiliation{ARC Centre of Excellence for Particle Physics at the Terascale, School of Physics, The University of Melbourne, Victoria 3010, Australia}

\begin{abstract}

We construct Lifshitz field theories in $4+1$ dimensions which retain $3+1$-d Lorentz invariance
and therefore ensure a unique limiting speed in the $3+1$-d world.
Such a construction is potentially useful in developing field-theoretic ultraviolet completions of
extra-dimensional field theories.  The extra dimension $y$ is treated asymmetrically from the usual three spatial
dimensions by introducing derivatives of order $2n$ with respect to $y$ in the action.  We show that 
$\lambda \phi^4$ theory becomes progressively less non-renormalisable by power counting as $n$ is increased.
This suggests that the non-local theory obtained in the $n \to \infty$ limit may be complete in the ultraviolet.

\end{abstract}

\maketitle

\section{Introduction}

The proposal by Ho\v{r}ava to cure the ultraviolet (UV) divergence problems of perturbative general relativity
through the introduction of terms in the action involving high-order spatial derivatives (Lifshitz theories) \cite{Horava}, 
has spurred similar studies to be made of other non-renormalisable field theories \cite{Iengoetal, TV, HST}.   
This is especially relevant for field-theoretic models involving extra dimensions of space \cite{TV, HST, PPT}.  
As the number of spatial dimensions is increased, the UV behaviour of 
Lorentz-invariant field theories
becomes more problematic.  For example, $\lambda \phi^4$ theory in $3+1$ dimensions is renormalisable with the
coupling constant $\lambda$ having zero mass dimension, while in $4+1$-d or higher renormalisability is lost because
$\lambda$ then has negative mass dimension, just like Newton's constant in general relativity.  
Field-theoretic models with full extra-dimensional Lorentz invariance 
are hence necessarily effective theories, defined with a UV cut-off $M$.  For energies above
$M$, one must hope that a suitable UV completion exists, with some string theory being the usual suspect.  It would be
interesting to have an alternative paradigm for the UV completion of extra-dimensional field theories.  The Lifshitz 
class of theories with anisotropic scale invariance in the UV provides a candidate for such a structure.

Most studies of Lifshitz theories in the high-energy physics literature have treated time and space
asymmetrically, while maintaining isotropic scaling within the spatial dimensions themselves.  
This is an explicit breaking of Lorentz invariance.  The hope is that one can construct a theory that flows in
the infrared
to a fixed point that respects Lorentz invariance, so special relativity would be an emergent phenomenon, true only at 
sufficiently low energies.  So far, however, it has proven impossible to produce a low-energy effective theory which
naturally produces a common limiting speed for all particle species.  What happens instead is that, in general,
photons, electrons, neutrinos and so on all have different limiting speeds, and fine-tuning of parameters must
be undertaken to bring the theory into agreement with experimental bounds on the violation of special 
relativity \cite{Iengoetal}.

This problem requires a solution before realistic Lifshitz theories (in the high-energy physics domain) can be constructed.
The purpose of this paper is to explore a different implementation of Lifshitz scalar field theory in the extra-dimensional
context so as to avoid this problem entirely.  The idea is simply to treat the extra dimension(s) of space 
asymmetrically from the three usual spatial dimensions, while maintaining the usual relativistic equivalence of time
with those three dimensions: higher-dimensional Lorentz invariance will be absent, but the
usual $3+1$-d Lorentz symmetry will be exact (see also Ref.~\cite{HST, PPT}).  
We shall introduce higher derivatives with respect to the extra
dimension(s) in the action in order to quell the bad UV behaviour of standard extra-dimensional field theories.
We recognise that such a construction will not deal with the UV problem of general relativity, but we think the
development of UV-complete gravity-free field theories in higher dimensions is an interesting exercise in itself.

To start as simply as possible in Sec.~\ref{nth-order}, we shall consider only one extra dimension of space, 
and stick initially to scalar field theory.  The
minimal goal is to allow $\lambda \phi^4$ theory to be power-counting renormalisable.  Our main result 
in Sec.~\ref{pwr-counting} will be that 
one approaches this situation as one raises the order $n$ of the derivative in the $(\partial^n \phi/\partial y^n)^2$ term
in the Lagrangian.  We shall demonstrate that the order of perturbation theory at which new counter terms have to be
introduced to absorb divergences can be made arbitrarily large by increasing $n$ accordingly.  This suggests
that the non-local $n \to \infty$ theory may be UV complete, though in the present paper we shall restrict the
analysis to the finite $n$ case.  Section \ref{gen-theory} then briefly discusses the extension to a general gauge
theory in $4+1$ dimensions, while Sec.~\ref{conc} is a conclusion.

\section{Lifshitz theory with an nth-order derivative}
\label{nth-order}

Our prototype theory begins with the action
\begin{equation}
S = \frac{1}{2} \int d^4x\, dy \left[ \partial^\mu \phi \, \partial_\mu \phi - \frac{1}{\Lambda^{2(n-1)}} \left( 
\frac{\partial^n \phi}{\partial y^n} \right)^2 \right]\;,
\label{eq:protoS}
\end{equation}
where letters from the middle of the Greek alphabet such as $\mu$ denote the usual $3+1$-d Lorentz indices,
with $y$ the extra dimension.  Standard $3+1$-d Lorentz invariance holds for $S$, but the field $\phi(x,y)$ depends
both on $y$ and the usual coordinates $x$.  The order $n$ of the derivative is left arbitrary, and the
parameter $\Lambda$ has the dimension of mass in terms of the \emph{usual} natural units, which must not be 
confused with the anisotropic scaling dimension concept to be introduced next.  We may use the terms
``usual mass units'' and ``scaling mass units'', respectively.

Perform the scaling 
\begin{equation}
x \to \kappa x,\qquad y \to \kappa^z y, \qquad \phi \to \kappa^a \phi,
\end{equation}
and observe that $S$ is invariant when
\begin{equation}
z = \frac{1}{n},\qquad a = - 1 - \frac{1}{2n}\;.
\end{equation}
The case of usual $4+1$-d relativistic scalar field theory, $n=1$, sees full isotropic scaling between time and
the four spatial coordinates, with $\phi$ scaling as $\kappa^{-3/2}$.  The scaling mass dimension is the negative
of the exponent, so we recover the usual result $[\phi]=3/2$ for the mass dimension of a scalar in $4+1$-d.
For $n >1$, anisotropic scaling holds, and the scaling dimensions of $y$ and $\phi$, 
\begin{equation}
[x] = -1,\qquad [y] = - \frac{1}{n}, \qquad [\phi] = 1 + \frac{1}{2n}\;,
\end{equation}
differ from their familiar relativistic values.

Our goal is to have $\lambda \phi^4$ theory renormalisable (actually, less non-renormalisable) in $4+1$-d.  
The rules of power-counting
renormalisability require us to use the (anisotropic) scaling mass dimension rather than the natural mass dimension
to determine the dimension of $\lambda$.  A trivial calculation shows that
\begin{equation}
[\lambda] = - \frac{1}{n}\;,
\end{equation}
so any finite value of $n$ requires $\lambda$ to have a negative mass dimension thus making the theory
non-renormalisable.  But in the $n \to \infty$ limit, $\lambda$ becomes dimensionless.  This suggests
that a power-counting renormalisable theory may be produced in that limit, provided such a 
non-local theory actually makes sense.\footnote{Formally taking $n$ to be negative produces a $\lambda$
with positive scaling dimension, which may suggest renormalisability.  Mathematically there is, of course, 
a continuation of the
$n$th derivative to non-integer and negative values, so it may not be ridiculous to contemplate such a theory.
Note that negative $n$ implies integration (the inverse of differentiation), so once again a non-local theory
is obtained.  It is tempting to conjecture that there might be a general connection between UV completeness of
the kind of Lifshitz theories being considered here and non-locality.}
For this paper, we shall restrict our attention to finite $n$ and prove the following:
although renormalisability does not hold for any finite $n$, there is a well-defined sense
in which the UV behaviour steadily improves as $n$ is increased, tending to full renormalisability as $n \to \infty$.  

Following the prescription of power-counting renormalisability, one may add relevant operators to the Lagrangian
in $S$. These operators will explicitly break the scaling symmetry, but they will not affect the UV behaviour
of the theory, which will remain dominated by the marginal operators in Eq.~(\ref{eq:protoS}).
The only derivative-free relevant operators, and their scaling dimensions, are
\begin{equation}
[\phi^2] = - 2, \qquad [\phi^3] = -1 + \frac{1}{2n}\;.
\end{equation}
So, the quadratic term is the ``mass term'' as usual, and the cubic term is renormalisable for all positive
integer $n$.  For terms containing derivatives, 
the rule is
\begin{equation}
a < \frac{1}{2} [ 2 - b + 2 ( 4 - b ) n ]\;,
\end{equation}
where $a$ is the number of $\partial/\partial y$ factors, and $b$ is the number of $\phi$ factors.  Hence, any $b=2$ operator with $a < 2n$ is relevant.  For $b=3$, any
$a \le n-1$ produces a relevant operator.  For $b \ge 4$, there are no solutions for positive $a$.\footnote{Note that
at finite $n$, renormalisable interaction terms can exist, since $b=3$ terms are allowed (indeed $\phi^3$ is always
one of them, even for $n=1$).  But our goal is to have quartic renormalisable interactions, because 
we want potentials that are bounded from below and because cubic terms are often
forbidden by some internal symmetry.}
It is easy to verify
that there are no relevant operators involving derivatives with respect to $x$.

For simplicity, we shall impose the discrete symmetry $\phi \to -\phi$ to remove the $b=3$ terms.  We shall also enforce
$y$-parity symmetry, invariance under $y \to -y$, to remove terms that are odd in $\partial/\partial y$.
The full action of our free-field theory, containing both the marginal and relevant terms is
\begin{equation}
S = \frac{1}{2} \int d^4x dy  \left[ \partial^\mu \phi \partial_\mu \phi - \sum_{\alpha=1}^{n} 
\frac{a_\alpha}{\Lambda^{2(\alpha-1)}} \left( \frac{\partial^\alpha \phi}{\partial y^\alpha} \right)^2 - m^2 \phi^2 \right]\;,
\label{action_full}
\end{equation}
where we are free to set $a_n=1$.

Note that in the above action, we have assumed that terms such as 
$\phi (\partial^{2\alpha} \phi / \partial y^{2 \alpha})$, 
$( \partial \phi / \partial y)(\partial^{2\alpha - 1} \phi / \partial y^{2\alpha -1})$, and so on,
may always be reduced to 
$(\partial^{\alpha} \phi/\partial y^{\alpha})^2$ by successive integrations-by-parts, dropping the ``surface'' 
terms each time.  This procedure may be questionable in the non-local $n \to \infty$ limit, where boundary conditions
with respect to $y$ may need special treatment.  As a general statement, the $n \to \infty$ limit requires careful
consideration, because for any finite $n$ the theory is local, whereas in the limit it is non-local.

\section{Power-counting renormalisation analysis}
\label{pwr-counting}

From Eq.~(\ref{action_full}), the propagator for the scalar field $\phi$ in momentum space reads
\begin{equation} \label{propagator_full}
 \Delta_\phi(k,k_y) =
 \frac{i}{k^2 - \sum_{\alpha=1}^{n} G_\alpha k_y^{2\alpha} -m^2 +i\epsilon}\;,
\end{equation}
where $G_\alpha \equiv a_\alpha/\Lambda^{2(\alpha-1)}$ which has scaling dimension 
$\dim{G_\alpha}=2(1-\alpha/n)$,
$k^2=k^\mu k_\mu$ is the usual scalar product in $3+1$-d spacetime, and
$k_y$ denotes the momentum component along the extra-dimension $y$.
Note that the scaling dimensions for $k$, $k_y$ and $\Delta_\phi$ are, respectively,
\begin{equation}
 \dim{k} = 1\;, \qquad
 \dim{k_y} = \frac{1}{n}\;, \qquad
 \dim{\Delta_\phi}=-2
\;.
\end{equation}
In the limit of large $k_y$, the $n$th term of the summation in (\ref{propagator_full}) dominates and we effectively have
\begin{equation}
 \Delta_\phi(k,k_y) \simeq \frac{i}{k^2 - G_n k_y^{2n} - m^2 + i \epsilon}
\;.
\end{equation}
The rapid decay of the propagator with increasing $k_y$ for large $n$ is the reason higher spatial derivatives
soften the UV behaviour.

Having noted the novel scaling properties above, which encode the UV properties of our theory,
the power-counting analysis proceeds in a fairly straightforward manner, 
mimicking the usual procedure for a $\phi^4$-theory in $3+1$-d \cite{Srednicki}. 
By going through this argument,
we will be able to see exactly how the parameter $n$ progressively allows the ``bad''
divergences to be relegated to higher and higher orders of perturbation theory as it is increased.

So, we now add a $\lambda \phi^4$ term into our action.
We start by noting that the scaling dimension of any diagram (made up of $\lambda \phi^4$ vertices only) 
with an arbitrary number of external lines, \LE, is given by the sum of the dimensions of its components, namely
\begin{align}
 \diag{\LE}
   &= \dim{d^4k\, dk_y} L + \dim{\Delta_\phi} \LP + \dim{\lambda} \VL \;,\\
   &= \left(4+\frac{1}{n}\right) L -2\LP -\frac{\VL}{n} \label{dim_diag_LE}
\;,
\end{align}
where $L$, $\LP$ and $\VL$ denote the number of loops, propagators and $\lambda\phi^4$ vertices, 
respectively.\footnote{Without loss of generality, we may assume these diagrams are all one-particle 
irreducible.} For convenience, we have opted not to show the 
dependence on $\LE$ in \diag{\LE} explicitly at this point.

Since the coupling constant $\lambda$ is not a function of $k$ and $k_y$, the $\dim{\lambda}\VL$ 
component of (\ref{dim_diag_LE}) drops out when performing the loop 
integrations, and hence the 
dimensionality of that integral (the ``superficial degree of divergence'') is
\begin{align}
 \Dsup &= \left(4+\frac{1}{n}\right) L -2\LP\;,\\
 &= \diag{\LE} +\frac{\VL}{n}\label{D_sup}
 \;.
\end{align}
Applying the handshaking lemma and Euler's formula from graph theory to our system, we can write down, respectively,
\begin{equation}
 4\VL + \LE = 2(\LE +\LP)\;,
\end{equation}
and
\begin{equation}
 L+\VL =\LP+1\;.
\end{equation}
Substituting these into the definition for \diag{\LE}, we can express (\ref{D_sup}) as a function of $\LE$, $n$ and
$\VL$ as per
\begin{equation}\label{D_sup_LE}
 \Dsup = \left(4+\frac{1}{n}\right) - \left(1+\frac{1}{2n}\right)\LE +\frac{\VL}{n}
 \;.
\end{equation}
Observe that for a given $\LE$, the superficial degree of divergence gets worse as you increase the number of 
vertices (and hence loops) in a diagram. In other words, the theory will require an endless stream of 
counterterms corresponding to each $\LE$ case for sufficiently large \VL.  But (\ref{D_sup_LE}) also 
shows that we can simply choose a value for $n$ that is large enough to render the theory 
(superficially) renormalisable up to a certain order of perturbation theory. Requiring that the superficial
degree of divergence be negative for $\LE > 4$, we obtain the requirement
\begin{equation}
n > \frac{1 - \LE/2 + \VL}{\LE - 4}\;.
\label{n-ineq}
\end{equation}
For a given order of perturbation theory $\VL$, satisfaction of this inequality guarantees that all
loop graphs apart from $\LE = 2$ and $4$ are convergent, and thus no counterterms are required.  
Let us consider a simple example: suppose we want no divergences (apart from the 
removable ones for $2$- and $4$-point
functions) to appear before 6th-order.  Then Eq.~(\ref{n-ineq}) tells us we must take $n > 2$ 
(the strongest constraint is obtained from the smallest allowed $\LE > 4$, which is $6$).

Of course,
the superficial degrees of divergence for 2-point and 4-point functions, given, respectively, by
\begin{equation}
2 + \frac{\VL}{n}\qquad {\rm and}\qquad \frac{\VL-1}{n}
\end{equation}
are always non-negative, so those divergences must be removed in the standard way.  Observe that at a given
perturbation theory order, these divergences are softened by increasing $n$.
In the limit $n\rightarrow \infty$, the dependence on $\VL$ disappears and (\ref{D_sup_LE}) 
reduces to the standard $\phi^4$-theory result of $4 -\LE$. 

Increasing $n$ provides a tangible benefit in that it permits one to take the UV cut-off to larger
values.  Thus, taking the strict $n \to \infty$ limit may not be necessary in practice.  To quantify this, consider
the radiative generation of the $\phi^6$ term, which suffers from the most severe irremovable divergence.  The
coefficient $C_6$ of this term has scaling dimension $-2(1+1/n)$, so we may write it as
\begin{equation}
C_6 = \frac{c}{U^{2 \left( 1 + \frac{1}{n} \right)}},
\end{equation}
where $c$ is dimensionless and $U$ is the UV cut-off with a scaling dimension equal to one.  At order $\VL$,
and taking $U \gg m$, this coefficient evaluates approximately to
\begin{equation}
C_6 \sim \tilde{\lambda}^{\VL} U^{\Dsup}
\end{equation}
where $\tilde{\lambda}$ is the quartic coupling constant $\lambda$ divided by a dimensionless loop factor.
The loop integral supplies the cut-off dependent factor $U^{\Dsup}$ as per the superficial degree of divergence.
Now, for $\LE = 6$,
\begin{equation}
\Dsup = - 2 \left( 1 + \frac{1}{n} \right) + \frac{\VL}{n}
\end{equation}
according to Eq.~(\ref{D_sup_LE}), so that
\begin{equation}
c \sim \left( \tilde{\lambda} U^{\frac{1}{n}} \right)^{\VL}.
\end{equation}
If the quantity in parentheses (which has scaling dimension of zero) is less than one, then the terms in the 
perturbation series decrease as $\VL$
increases, which is what we need in order for perturbation theory to make sense.  This puts an upper bound on the 
UV cut-off: $U \lesssim (1/\tilde{\lambda})^n$.  For a given value of $\tilde{\lambda}$, we see that higher
values for $U$ are permitted as $n$ is increased.

\section{Extension to general gauge theory}
\label{gen-theory}

A simple examination of Yukawa and gauge extensions of the above suggests that a general gauge theory
of scalars and fermions shares the feature of becoming less non-renormalisable as $n$ is increased, though a
thorough study of this is beyond the scope of this paper.

The marginal $3+1$-d Lorentz invariant kinetic energy operators for a Dirac field $\psi(x,y)$ provide the action
\begin{equation}
S_\psi = \int d^4x\, dy \left[ i \bar{\psi} \gamma^\mu \partial_{\mu} \psi + \bar{\psi} (a + b \gamma_5) 
\frac{\partial^n \psi}{\partial y^n} \right],
\label{eq:fermion}
\end{equation}
where $a (b) $ is real (pure imaginary) for even $n$ and pure imaginary (real) for odd $n$ to ensure hermiticity.
This action tells us that the scaling dimension of the Dirac field is
\begin{equation}
[ \psi ] = \frac{3}{2} + \frac{1}{2n}.
\end{equation}
This means that Yukawa coupling terms $h_1 \bar{\psi} \psi \phi + h_2 \bar{\psi} \gamma_5 \psi \phi$ feature 
Yukawa coupling constants with dimension
\begin{equation}
[ h_{1,2} ] = - \frac{1}{2n},
\end{equation}
so they share the property with quartic coupling constants of becoming dimensionless as $n$ tends to infinity.

The standard gauge field kinetic term $-(1/4) {\rm Tr} F^{\mu\nu} F_{\mu\nu}$, where the trace is over
suitably normalised internal symmetry generators, enforces
\begin{equation}
[ A ] = 1 + \frac{1}{2n}
\end{equation}
for the gauge field $A^\mu(x,y)$. The scaling dimension of a gauge coupling constant $g$, defined
by the covariant $3+1$-d derivative $\partial_\mu - i g A_{\mu}$, is then obviously
\begin{equation}
[ g ] = - \frac{1}{2n},
\end{equation}
just as for the Yukawa coupling constant.  Since both gauge and Yukawa coupling constants become 
dimensionless as $n$ tends to infinity, it is probable that a general Lifshitz gauge theory of this type 
also has the property of becoming less non-renormalisable as $n$ increases.

There are two \emph{a priori} possible gauge structures.  One is based on the observation that
the restriction to $3+1$-d Lorentz invariance suggests that the gauge field need only have four components,
with the gauge transformation
\begin{equation}
A_\mu(x,y) \to U(x) A_\mu(x,y) U(x)^{\dagger} - \frac{i}{g} (\partial_{\mu}U(x)) U(x)^{\dagger}
\end{equation}
depending only on $x$, not on both $x$ and $y$.  In this case, the 
gauge-invariant Lifshitz term ${\rm Tr} (\partial^n A^\mu / \partial y^n)(\partial^n A_\mu / \partial y^n)$ can be
introduced.  Gauge invariance is essential in producing standard renormalisable theories of vector fields, 
so one should be cautious about restricting the gauge functions $U$ to only being dependent on $x$,
which motivates the second gauge structure.  If the gauge
transformations are required to depend on $y$ as well as $x$, then the above Lifshitz term is forbidden and a 
fifth component $A_5$ must be introduced to produce a covariant derivative
$D_y \equiv \partial/\partial y - i g A_5(x,y)$ along the $y$-direction. 
The gauge-invariant term $-(1/2){\rm Tr} F^{\mu 5} F_{\mu 5}$ then provides the kinetic and gradient energy for $A_5$.  
Since this term
necessarily contains $x$-derivatives, higher powers of it are disallowed by the need to avoid the higher
time derivatives that would spoil unitarity.  Note that it is a relevant operator.  This kind of Lifshitz gauge theory
was recently studied in Ref.~\cite{HST}.  Note that the nth ordinary derivative in Eq.~(\ref{eq:fermion}) would need to be
replaced by the nth power of $D_y$.  Both of the above gauge structures feature a $[ g ] = - 1/2n$ gauge coupling
constant, so the fact that it becomes dimensionless in the infinite $n$ limit is independent of which structure is
ultimately seen to be superior.

\section{Conclusion}
\label{conc}

We have introduced a new kind of Lifshitz theory that may have uses in field-theoretic models involving extra
dimensions of space.  It maintains $3+1$-d Lorentz invariance, thus ensuring a common limiting speed on that
submanifold, but the extra dimension is treated differently.  Derivatives of arbitrarily high order with respect to the
extra dimension are introduced in order to suppress the propagation of states which have large momentum 
components along that direction.  For $\lambda \phi^4$ theory in $4+1$-d, we proved that the theory becomes less 
non-renormalisable as the order of the derivative is increased, and suggested that this is also true for a general
gauge theory of scalars and fermions including Yukawa coupling terms.  We speculate that the non-local theory
obtained as the order of the derivative goes to infinity may be ultra-violet complete.  The suppression of momentum
components along the extra dimension also suggests a connection with dynamical localisation and dimensional
reduction.

\section*{Acknowledgements}

We thank A. Kobakhidze and A. Zee for comments on the manuscript.
This work was supported in part by the Australian Research Council. XGH is partially supported by the NSC 
and NCTS of Taiwan, and SJTU 985 grant of China. SSCL is supported in part by the NSC and in 
part by the NCTS of Taiwan.

\end{document}